\documentstyle[12pt]{article}
\def\thebibliography#1{\bigskip\section*{\centering
References\\}\bigskip\list
{\arabic{enumi}.}{\settowidth\labelwidth{#1}\leftmargin\labelwidth
\advance\leftmargin\labelsep
\usecounter{enumi}}
\def\newblock{\hskip .11em plus .33em minus .07em}
\sloppy\clubpenalty4000\widowpenalty4000
\sfcode`\.=1000\relax}

\def\op#1{\mathop{\fam0 #1}\limits}

\newcommand{\ben}{\begin{eqnarray}}
\newcommand{\een}{\end{eqnarray}}

\newcommand{\be}{\begin{eqnarray*}}
\newcommand{\ee}{\end{eqnarray*}}
\newcommand{\bea}{\begin{eqalph}}
\newcommand{\eea}{\end{eqalph}}
\newcommand{\cL}{{\cal L}}
\newcommand{\bL}{{\bf L}}

\newcommand{\cE}{{\cal E}}

\newcommand{\cF}{{\cal F}}

\newcommand{\al}{\alpha}
\newcommand{\bt}{\beta}
\newcommand{\kp}{\kappa}
\newcommand{\dl}{\delta}
\newcommand{\la}{\lambda}

\newcommand{\ap}{\approx}

\newcommand{\om}{\omega}
\newcommand{\ot}{\otimes}
\newcommand{\Om}{\Omega}
\newcommand{\m}{\mu}
\newcommand{\n}{\nu}
\newcommand{\g}{\gamma}

\newcommand{\ve}{\varepsilon}

\newcommand{\Si}{\Sigma}
\newcommand{\si}{\sigma}

\newcommand{\w}{\wedge}

\newcommand{\wt}{\widetilde}
\newcommand{\wh}{\widehat}
\newcommand{\ol}{\overline}
\newcommand{\dr}{\partial}

\newcounter{eqalph}
\newcounter{equationa}

\newenvironment{eqalph}{\stepcounter{equation}
\setcounter{equationa}{\value{equation}}
\setcounter{equation}{0}

\begin{eqnarray}}{\end{eqnarray}
\setcounter{equation}{\value{equationa}}}

\begin{document}
\hbox{}

\centerline{\large\bf Stress-Energy-Momentum of Affine-Metric Gravity.}
\medskip

\centerline{\large\bf Generalized Komar Superpotential.}
\bigskip

\centerline{\sc Giovanni Giachetta}

\medskip

\centerline{Department of Mathematics and Physics}

\centerline{University of Camerino, 62032 Camerino, Italy}

\centerline{E-mail: mangiarotti@camvax.unicam.it}
\medskip

\centerline{\sc Gennadi Sardanashvily}
\medskip

\centerline{Department of Theoretical Physics}

\centerline{Moscow State University, 117234 Moscow, Russia}

\centerline{E-mail: sard@grav.phys.msu.su}
\bigskip

\begin{abstract}

In case of the Einstein's gravitation theory and its first order Palatini
reformulation, the stress-energy-momentum of gravity has been proved to
reduce to the Komar superpotential. We generalize this result to the
affine-metric theory of gravity in case of general connections and arbitrary
Lagrangian densities invariant under general covariant transformations. In
this case, the stress-energy-momentum of gravity comes to the generalized Komar
superpotential depending on a Lagrangian density in a precise way.
\end{abstract}

\section{}

As  is well-known, in the Einstein's gravitation theory of
metric fields \cite{nov}, the stress-energy-momentum (SEM) conservation law
 corresponding to the invariance of the Hilbert-Einstein
Lagrangian density under general covariant transformations takes the form
\begin{equation}
\frac{d}{dx^\la}U(\tau)^\la = 0 \label{K1}
\end{equation}
where
\begin{equation}
U(\tau)^\la =\frac{d}{dx^\m} [\frac{\sqrt{-g}}{2\kp}
( g^{\la\nu}\tau^\m_{;\nu} -g^{\m\nu}\tau^\la_{;\nu})]  \label{K2}
\end{equation}
is the well-known Komar superpotential \cite{kom} associated with a vector
field $\tau$ on a world manifold $X$. By the symbol "$_{;\m}$" is meant the
covariant derivative with respect to the Levi-Civita connection.

In a recent paper \cite{bor}, it was shown that, in the Palatini model of
metric fields and symmetric connections , the SEM complex corresponding to a
Lagrangian density polinomial in a scalar curvature looks like the Komar
superpotential (\ref{K2}).

We generalize this result to the affine-metric gravity in case of a general
linear connection $K^\al{}_{\g\m}$ and arbitrary Lagrangian density $L$
invariant under general covariant transformations. The corresponding
SEM conservation law  \cite{cam1} is brought into the form (\ref{K1}) where
\begin{equation}
U(\tau)^\la =\frac{d}{dx^\m} [\frac{\dr\cL}{\dr K^\al{}_{\nu\m,\la}} (D_\nu
\tau^\al + \Om^\al{}_{\nu\si}\tau^\si)]   \label{K3}
\end{equation}
is the generalized Komar superpotential (\ref{K2}). Here, by $D_\g$ is
meant the covariant derivative with respect to the general linear connection
$K$ and $\Om$ is the torsion of this connection. In the particular case of
the Hilbert-Einstein Lagrangian density and symmetric connections, we have
$$
\frac{\dr\cL_{\rm HE}}{\dr K^\al{}_{\nu\m,\la}} =
\frac{\sqrt{-g}}{2\kp}
(\dl^\m_\al g^{\nu\la} - \dl^\la_\al g^{\nu\m}),
$$
so that the superpotential (\ref{K3}) comes to the standard Komar
superpotential (\ref{K2}).
If a Lagrangian density is a polynomial in the scalar curvature of a
symmetric connection, the superpotential (\ref{K3}) recovers the one in ref.
\cite{bor}.

\section{}

We follow the geometric approach to field theory when classical fields are
described by global sections of a bundle $Y\to X$ over a world manifold $X$.
Their dynamics is phrased in terms of jet manifolds
\cite{sard,sard0,sau}.

As a shorthand, one can say that the
$k$-order jet manifold $J^kY$ of a bundle $Y\to X$
comprises the equivalence classes
$j^k_xs$, $x\in X$, of sections $s$ of $Y$ identified by the first $k+1$
terms of their Taylor series at a point $x$.  Recall that
a $k$-order differential operator on sections of a bundle $Y\to X$ is defined
to be a bundle morphism of the bundle $J^kY\to X$ to a vector
bundle over $X$.

We restrict ourselves
to the first order  Lagrangian formalism, for
 most of contemporary field models are described by
first order Lagrangian densities. This is not the case for
 the Einstein-Hilbert Lagrangian density of the Einstein's
gravitation theory which belongs to the special class of second order
Lagrangian densities whose Euler-Lagrange equations are however of the order
two as like as in the first order theory.

In the first order Lagrangian formalism, the finite-dimensional
configuration space
of fields represented by sections $s$ of a bundle $Y\to X$ is the first order
jet manifold $J^1Y$ of $Y$.  Given fibered coordinates $(x^\m,y^i)$
of $Y$, the jet manifold $J^1Y$ is endowed with the adapted  coordinates
$ (x^\m,y^i,y^i_\m)$. In physical literature, the coordinates $y^i_\m$ are
usually called the velocity coordinates or the derivative coordinates
because of the relation
$$
y^i_\m(j^1_x s) =\dr_\m s^i(x).
$$
The jet manifold $J^1Y$ is endowed with the natural bundle structures
$J^1Y\to Y$ and $J^1Y\to X$. For the sake of convenience, we shall call
$J^1Y\to X$ the configuration bundle and $J^1Y\to Y$ simply the jet
bundle.

A first order
Lagrangian density on $J^1Y$ is defined to be an exterior horizontal density
\be
&& L: J^1Y\to\op\w^nT^*X, \qquad n=\dim X,\\
&&L=\cL(x^\m,y^i,y^i_\m)\om, \qquad \om=dx^1\w ...\w dx^n,
\ee
on the configuration bundle $J^1Y\to X$.

By a differential
conservation law in first order field theories is meant a relation where
the divergence of a current $T$
appears equal to zero, i.e.
\begin{equation}
dT =0 \label{C1}
\end{equation}
where $T$ is a horizontal $(n-1)$-form on the configuration bundle $J^1Y\to
X$.

The relation (\ref{C1}) is called a strong conservation law if it
is satisfied identically by all sections $s$ of the bundle $Y\to X$, while it
is termed a weak conservation law if it takes place only on
critical sections, i.e.,  on solutions of field equations. We shall use the
symbol "$\ap$" for weak identities.

It may happen that
the weakly conserved current $T$ takes the form
$$
T=W+dU
$$
where $W\ap 0$. In this case,
one says that the current $T$ is reduced to the superpotential $U$
\cite{bor,fat}. For instance, the N\"other currents in gauge theory come to
superpotentials which depend on parameters of gauge transformations that
provide the gauge invariance of N\"oether conservation laws.

Usually, one derives the differential conservation
laws from invariance of a Lagrangian density under some group of
transformations.

Let $G_t$ be a 1-parameter group of bundle isomorphisms of a bundle $Y\to X$
and
$$
u= u^\la (x)\dr_\la + u^i(y)\dr_i
$$
the corresponding vector field on $Y$. One can prove that a
Lagrangian density  $L$ on the configuration space $J^1Y$ is invariant under
these transformations iff its Lie derivative by the jet lift
$$
j^1_0u=u^\la\dr_\la + u^i\dr_i + (\wh\dr_\la u^i
- y_\m^i\dr_\la u^\m)\dr_i^\la,
$$
$$
\wh\dr_\la = \dr_\la +y^j_\la \dr_j+y^j_{\la\m}\dr_j^{\la\m} +\cdots
$$
of $u$ onto $J^1Y$ is equal to zero:
\begin{equation}
\bL_{j^1_0u}L=0. \label{C2}
\end{equation}
The equality (\ref{C2}) gives rise to
the weak differential conservation law as follows.

Let $u$
be a projectable vector field on a bundle $Y\to X$ and $\ol u$ its jet lift
 onto the configuration space $J^1Y\to X$. Given a
Lagrangian density   $L$, let us compute the Lie derivative ${\bf L}_{\ol
u}L$. We get the canonical decomposition
\begin{equation}
\bL_{j^1_0u}L=  u_V\rfloor\cE_L+ dT(u) \label{C3}
\end{equation}
where
$$
\cE_L=
 [\dr_i-(\dr_\la +y^j_\la\dr_j+y^j_{\m\la}\dr^\m_j)
\dr^\la_i]\cL dy^i\w\om=\cE_idy^i\w\om
$$
is the Euler-Lagrange operator,
$$
T(u)=T^\la(u)\om_\la =
\wh \dr_\la[\pi^\la_i(u^i-u^\m y^i_\m )+u^\la\cL]\om_\la,
$$
$$
\wh\dr_\la =\dr_\la +y^i_\la\dr_i+y^i_{\m\la}\dr^\m_i,
\qquad \om_\la =\dr_\la\rfloor\om,
$$
is the corresponding current,
and
$$
u_V=(u^i -y^i_\m u^\m)\dr_i
$$
is the vertical part of the vector field $u$.
We denote by $\pi^\m_i=\dr^\m_i\cL$
the Lagrangian momenta.
This is
the well-known first variational formula of the calculus of
variations.

The
Euler-Lagrange operator  $\cE_L$, by definition, vanishes on the critical
sections of the bundle $Y\to X$, and the equality (\ref{C3}) comes to the weak
identity
$$
 {\bf L}_{j^1_0u}L\ap \wh\dr_\la[\pi^\la_i(u^i- u^\m
y^i_\m) +u^\la\cL ]\om.
$$

If the Lie derivative
$$
\bL_{\ol u}L=[\dr_\la u^\la\cL +(u^\la\dr_\la+
u^i\dr_i +(\dr_\la u^i +y^j_\la\dr_ju^i -y^i_\m\dr_\la u^\m)\dr^\la_i)\cL]\om
$$
of a Lagrangian density $L$ by a projectable vector field $u$ satisfies the
condition (\ref{C2}), then
we have the weak conservation law
\begin{equation}
0\ap \wh \dr_\la[\pi^\la_i(u^i-u^\m y^i_\m )+u^\la\cL]. \label{K4}
\end{equation}

There are two theories where the condition (\ref{C2}) takes place:
(i) the gauge theory of exact internal symmetries where $u$ are vertical fields
associated with gauge transformations and (ii) the gravitation theory on
bundles of geometric objects $Y\to X$ where $u$ are canonical lifts of vector
fields $\tau$ on the base $X$ onto $Y$. In this case, we have
the SEM conservation laws \cite{cam1}.

\section{}

The bundles of geometric objects $Y\to X$ are examplified by tensor bundles
over $X$ and the bundle of linear connections. They admit the canonical lift
of a vector field $\tau$ on $X$.

Let $\tau=\tau^\m\dr_\m$ be a vector field on the manifold $X$. There exists
the canonical lift
\begin{equation}
\wt\tau =T\tau= \tau^\m\dr_\m +\dr_\nu\tau^\al\dot x^\nu\frac{\dr}{\dr\dot
x^\al} \label{C91}
 \end{equation}
of $\tau$ onto the tangent bundle $TX$ of $X$.
This lift
consists with the horizontal lift of $\tau$ by means the symmetric connection
$K$ on the tangent bundle which has $\tau$
as the geodesic field:
$$
\dr_\nu\tau^\al +K^\al{}_{\m\nu}\tau^\m=0.
$$

Generalizing the canonical
lift (\ref{C91}), one can construct the lifts of a vector field
$\tau$ on $X$ onto the following bundles over $X$ (for the sake of simplicity,
we denote all these lifts by the same symbol $\wt\tau$):
We have:
\begin{itemize}
\item the canonical lift
$$
\wt\tau = \tau^\m\dr_\m -\dr_\bt\tau^\nu\dot x_\nu\frac{\dr}{\dr\dot x_\bt}
$$
of $\tau$ onto the cotangent bundle $T^*X$;

\item the canonical lift
$$
\wt\tau = \tau^\m\dr_\m + [\dr_\nu\tau^{\al_1}
\dot x^{\nu\al_2\cdots\al_m}_{\bt_1\cdots\bt_k} + \ldots -
\dr_{\bt_1}\tau^\nu \dot x^{\al_1\cdots\al_m}_{\nu\bt_2\cdots\bt_k} -\ldots]
\frac{\dr}{\dr \dot x^{\al_1\cdots\al_m}_{\bt_1\cdots\bt_k}}
$$
of $\tau$ onto the tensor bundle
$$
T^k_mX=(\op\ot^mTX)\ot(\op\ot^kT^*X);
$$

\item the canonical lift
$$
\wt\tau = \tau^\m\dr_\m +[\dr_\nu\tau^\al k^\nu{}_{\bt\m} - \dr_\bt\tau^\nu
k^\al{}_{\nu\m} - \dr_\m\tau^\nu
k^\al{}_{\bt\nu} -\dr_{\bt\m}\tau^\al]\frac{\dr}{\dr k^\al{}_{\bt\m}}
 $$
of $\tau$ onto the bundle $C$ of the linear connections on $TX$.
\end{itemize}

One can think of the vector fields  $\wt\tau$ on a bundle of geometric objects
$Y$ as being the vector fields associated with the local 1-parameter groups of
the holonomic isomorphisms of $Y$ induced by diffeomorphisms of its base $X$.
In particular, if $Y=TX$ they are the tangent isomorphisms. We call these
isomorphisms the general covariant transformations.

Let $Y$ be the bundle of geometric objects and $L$ a Lagrangian density on
the configuration space $J^1Y$. Given a vector field $\tau$ on the base $X$
and its canonical lift $\wt\tau$ onto $Y$, one may utilize the first
variational formula (\ref{C3}) in order to get the corresponding SEM
transformation law. If the Lagrangian density $L$ is invariant under general
covariant transformations, we have the equality
\begin{equation}
\bL_{j^1_0\wt\tau}L=0 \label{C95}
\end{equation}
and get the weak conservation law (\ref{K4}).
One can show that the conserved quantity is reduced to a superpotential term.

Let us verify this fact in case of a tensor bundle $Y\to X$. Let it be
coordinatized by $(x^\la, y^i)$ where the collective index $A$ is employed.
Given a vector field $\tau$ on $X$, its canonical lift $\wt\tau$ on $Y$ reads
$$
\wt\tau =\tau^\la\dr_\la + u^i{}_\al^\bt\dr_\bt\tau^\al\dr_i.
$$

Let a  Lagrangian density $L$ on the configuration space $J^1Y$ be
invariant under general covariant transformations. Then, it satisfies the
equality (\ref{C95}) which takes the coordinate
form
\begin{equation}
\dr_\al(\tau^\al\cL) + u^i{}_\al^\bt\dr_\bt\tau^\al\dr_i\cL +
\wh\dr_\m(u^i{}_\al^\bt\dr_\bt\tau^\al)\dr_i^\m\cL -
 y^i_\al\dr_\bt\tau^\al\dr_i^\bt\cL =0. \label{C310}
\end{equation}
Due to the arbitrariness of the functions $\tau^\al$, the equality
(\ref{C310}) is equivalent to the system of equalities
\bea
 && \dr_\la\cL=0, \\
&& \dl^\bt_\al\cL + u^i{}_\al^\bt\dr_i\cL +
\wh\dr_\m(u^i{}_\al^\bt)\dr_i^\m\cL -
 y^i_\al\dr_i^\bt\cL =0,\label{C311a}\\
&& u^i{}_\al^\bt\dr_i^\m\cL +u^i{}_\al^\m\dr_i^\bt\cL =0. \label{C311b}
\eea
It is readily observed that the equality (\ref{C311a}) can be brought into
the form
\begin{equation}
\dl^\bt_\al\cL + u^i{}_\al^\bt\cE_i +
\wh\dr_\m(u^i{}_\al^\bt\dr_i^\m\cL) =
 y^i_\al\dr_i^\bt\cL\label{C312}
\end{equation}
where $\cL_i$ are the variational derivatives of the Lagrangian density
$L$. Substituting the relations (\ref{C312}) and (\ref{C311b})
into the weak identity
$$
0\ap \wh\dr_\la [(u^i{}_\al^\bt\dr_\bt\tau^\al -y^i_\al\tau^\al)\dr^\la_i\cL
+\tau^\la\cL]
$$
(\ref{K4}), we get the conservation law
\begin{equation}
0\ap \wh\dr_\la [-u^i{}_\al^\la \cE_i\tau^\al -
\wh\dr_\m(u^i{}_\al^\la\dr_i^\m\cL\tau^\al)]
\label{C313}
\end{equation}
where the conserved current is reduced to the superpotential term
\begin{equation}
Q(\wt\tau)^\la = -u^i{}_\al^\la\cL_i\tau^\al -
\wh\dr_\m(u^i{}_\al^\la\dr_i^\m\cL\tau^\al).
\label{C314}
\end{equation}

One can utilize the tensor fields described above as a matter source in the
gravitation theory on bundles of geometric objects. Their Lagrangian
densities are independent on a symmetric connection, and they contain the
torsion of a general linear connection.

\section{}

Proca fields which are described by sections of the
cotangent bundle $Y=T^*X$ examplify a field model on bundles of geometric
objects. Let us consider
the SEM transformation law of Proca fields in the presence of a background
world metric.

The
configuration space
$J^1Y$
of Proca fields is coordinatized by
$(x^\la,k_\mu,k_{\mu\la})$
where $k_\mu=\dot x_\mu$ are the induced coordinates of $T^*X$.
The Lagrangian density of Proca fields is
\begin{equation}
L_{\rm P}=[-\frac{1}{4\g}g^{\mu\al}g^{\nu\beta}\cF_{\al
\beta}\cF_{\mu\nu}
 -\frac12 m^2g^{\mu\la}k_\mu k_\la]\sqrt{\mid g\mid}\omega \label{214}
\end{equation}
where
$$
\cF_{\mu\nu}=k_{\nu\m} -k_{\mu\nu}.
$$

Let $\tau$ be a vector field on the base $X$ and
$$
\wt\tau = \tau^\m\dr_\m -\dr_\al\tau^\nu k_\nu\frac{\dr}{\dr k_\al}
$$
its canonical lift onto $T^*X$.

The Lie
derivative of the Lagrangian density (\ref{214}) by the jet lift
$j^1_0\wt\tau$
of the field $\wt\tau$ is
$$
\bL_{j^1_0\wt\tau}L_{\rm P}=(\dr_\la\tau^\la \cL_{\rm P}+\tau^\la\dr_\la
\cL_{\rm P}-
\cF_{\m\nu}\dr_\la\tau^\m\pi^{\nu\la} + m^2g^{\nu\la}\dr_\la\tau^\m
k_\nu k_\m\sqrt{\mid g\mid})\om
 $$
where $\pi^{\nu\la}=\dr^{\nu\la}\cL$ are the Lagrangian momenta.
Then, the corresponding SEM transformation law
$$
\bL_{j^1_0\wt\tau}L_{\rm P}\ap \wh\dr_\la[\pi^{\nu\la}(-\dr_\nu\tau^\m k_\m
-\tau^\m k_{\nu\m}) +\tau^\la\cL]
 $$
takes the form
\begin{equation}
-\dr_\la \tau^\m t_\m^\la\sqrt{\mid g\mid}-\tau^\m t_\bt^\al\sqrt{\mid
g\mid}\{^\bt{}_{\m\al}\} \ap
 \wh\dr_\la[-\tau^\m t_\m^\la\sqrt{\mid g\mid}+ \tau^\nu k_\nu\cE^\la
-\wh\dr_\m(\pi^{\mu\la}\tau^\nu k_\nu)] \label{C305}
\end{equation}
where
$$
t_\mu^\la\sqrt{\mid g\mid}= 2g^{\la\nu}\frac{\dr\cL}{\dr g^{\nu\mu}}
$$
is the energy-momentum tensor.

A glance at the expression (\ref{C305}) shows that the SEM tensor of the
Proca field
\begin{equation}
T(\ol\tau)^\la =\tau^\m t_\m^\la\sqrt{\mid g\mid}- \tau^\nu
k_\nu\cE^\la +\wh\dr_\m(\pi^{\mu\la}\tau^\nu k_\nu) \label{307}
\end{equation}
is the sum of the familiar metric energy-momentum tensor
 and the superpotential term
\begin{equation}
Q(\ol\tau)^\la = - \tau^\nu k_\nu\cE^\la
+\wh\dr_\m(\pi^{\mu\la}\tau^\nu k_\nu) \label{C306}
\end{equation}
which is the particular case of the superpotential term (\ref{C314}).
This term however does not make any contribution into the differential
conservation law (\ref{C305}) which thus takes the standard form
$$
 t^\la_{\m;\la}\ap 0.
$$

At the same time, the superpotential term reflects the fact that the symmetry
of the Lagrangian density (\ref{214}) under general covariant transformations
is broken by the background metric field. In gravitation theory, when
the general covariant transformations are exact,  the total superpotential
term contains
the whole SEM tensor
(\ref{307}) of Proca fields. Thus,
the Proca field model examplifies the phenomenon of "hidden energy". Only the
superpotential part of energy-momentum is displayed if the general covariant
transformations are exact.

\section{}

Let us consider the affine-metric gravitational model where dynamic variables
are  pseudo-Riemannian metrics and general
linear connections on $X$. They are called the
world metrics and the world connections respectively.

The 4-dimensional base manifold $X$ is assumed to satisfy the well-known
topological condition in order that it can be provided with a
pseudo-Riemannian  metric.  We call it  the world manifold.

Let $LX\to X$ be the principal bundle of linear frames in the
tangent spaces to $X$. Its structure group is $GL^+(4,{\bf R})$.

World metrics are represented by sections of the bundle
$\Si_g\to X$. It is the
2-fold covering of the bundle
$$
\Si=LX/SO(3,1)
$$
where $SO(3,1)$ is the
connected Lorentz group. Hereafter, we shall identify $\Si_g$ with the open
subbundle of the tensor bundle
$$
\op\vee^2T^*X\to X.
$$
The bundle $\Si_g$ is coordinatized by
$(x^\la, g_{\al\bt})$.

The world connections are thought as principal connections on the
principal bundle $LX\to X^4$.
 Indeed, there is the 1:1 correspondence between the
world connections and the global sections of the bundle
$$
C=J^1LX/GL^+(4,{\bf R}).
$$
With
respect to a holonomic atlas,  the bundle $C$ is coordinatized by
$(x^\la, k^\al{}_{\bt\la})$ so that, for any section $K$ of $C$,
$$
K^\al{}_{\bt\la}= k^\al{}_{\bt\la}\circ K
$$
are the coefficients of the linear connection
$$
K=dx^\la\otimes (\frac{\dr}{\dr x^\la} + K^\al{}_{\bt\la}\dot x_\al
\frac{\dr}{\dr \dot x^\bt})
$$
on $T^*X$.

Note that, since the world
connections are the principal connections,
one may apply the standard procedure of
gauge theory, but in this case the
nonholonomic gauge isomorphisms of the linear frame bundle $LX$ and
the associated bundles to be considered \cite{heh}. The canonical lift
$\ol\tau$ of a vector field $\tau$ on the base $X$ onto the bundles $\Si_g$
and $C$ does not correspond to these isomorphisms. One must use a horizontal
lift of $\tau$ by means of some connections on these bundles. Here, we limit
our consideration to the framework of field theory on the bundles of geometric
objects.

The total configuration space of the
affine-metric gravity is
\begin{equation}
J^1Y=J^1(\Si_g\op\times_XC) \label{N33}
\end{equation}
coordinatized by
$$
(x^\la, g^{\al\bt}, k^\al{}_{\bt\la}, g^{\al\bt}{}_\m,
k^\al{}_{\bt\la\m}).
$$

We assume that a Lagrangian density $L$ of the affine-metric gravitation theory
on the configuration space (\ref{N33}) depends on a metric $g^{\al\bt}$ and the
curvature
\[
R^\al{}_{\bt\n\la}=k^\al{}_{\bt\la\n}-
k^\al{}_{\bt\n\la}+k^\al{}_{\ve\n}k^\ve{}_{\bt\la}-k^\al{}_{\ve\la}
k^\ve{}_{\bt\n}.
\]
In this case, we have the relations
\be
&&\frac{\dr\cL}{k^\al{}_{\bt\nu}}= \pi_\si{}^{\bt\nu\la}k^\si{}_{\al\la} -
\pi_\al{}^{\si\nu\la}k^\bt{}_{\si\la},\\
&& \pi_\al{}^{\bt\nu\la}=\dr_\al{}^{\bt\nu\la}\cL =
-\pi_\al{}^{\bt\la\nu}.
\ee

Let the Lagrangian density $L$ be invariant under general covariant
transformations.
Given a vector field $\tau$ on $X$, its canonical lift onto the bundle
$\Si_g\times C$ reads
\ben
&&\wt\tau =\tau^\la\dr_\la + (g^{\nu\bt}\dr_\nu\tau^\al
+g^{\al\nu}\dr_\nu\tau^\bt)\frac{\dr}{\dr g^{\al\bt}}\nonumber\\
&& \qquad +[\dr_\nu\tau^\al k^\nu{}_{\bt\m} - \dr_\bt\tau^\nu
k^\al{}_{\nu\m} - \dr_\m\tau^\nu
k^\al{}_{\bt\nu} -\dr_{\bt\m}\tau^\al]\frac{\dr}{\dr k^\al{}_{\bt\m}}.
\label{C151}
\een
For the sake of simplicity, the compact notation
$$
\wt\tau =\tau^\la\dr_\la + (g^{\nu\bt}\dr_\nu\tau^\al
+g^{\al\nu}\dr_\nu\tau^\bt)\dr_{\al\bt} + (u^A{}_\al^\bt\dr_\bt\tau^\al
-u^A{}_\al^{\ve\bt}\dr_{\ve\bt}\tau^\al)\dr_A
$$
is employed.

Since the Lie derivative of $L$
by the jet lift
$j^1_0\wt\tau$ of the field $\wt\tau$ (\ref{C151})
is equal to zero:
\begin{equation}
\bL_{j^1_0\wt\tau}L=0, \label{K6}
\end{equation}
we have the weak conservation law
\begin{equation}
0\ap \wh\dr_\la[ \dr^\la_A\cL(u^A{}_\al^\bt\dr_\bt\tau^\al
-u^A{}_\al^{\ve\bt}\dr_{\ve\bt}\tau^\al -y^A_\al\tau^\la) +\tau^\la\cL]
\label{K8}
\end{equation}
where
\be
&& \dr^\la_A\cL u^A{}_\al^{\ve\bt}
=\pi_\al{}^{\ve\bt\la},\\
&& \dr^\ve_A\cL u^A{}_\al^\bt = \pi_\al{}^{\g\mu\ve}k^\bt{}_{\g\m} -
 \pi_\si{}^{\bt\mu\ve}k^\si{}_{\al\m} - \pi_\si{}^{\g\bt\ve}k^\si{}_{\g\al} =
\dr_\al{}^{\bt\ve}\cL - \pi_\si{}^{\g\bt\ve}k^\si{}_{\g\al}.
\ee

Due to the arbitrariness of the functions $\tau^\al$, (\ref{K6}) implies the
following equality
\begin{equation}
 \dl^\bt_\al\cL + \sqrt{-g}T^\al_\bt + u^A{}_\al^\bt\dr_A\cL +
\wh\dr_\m(u^A{}_\al^\bt)\dr_A^\m\cL -
 y^A_\al\dr_A^\bt\cL =0.\label{K9}
\end{equation}
One can think of
$$
\sqrt{-g}T^\al_\bt =2g^{\al\nu}\dr_{\nu\bt}\cL
$$
as being the metric energy-momentum tensor of general linear connections.

Substituting the term $y^A_\al\dr_A^\bt\cL$ from the expression (\ref{K9})
into the conservation law (\ref{K8}), we bring it into the form
\begin{equation}
0\ap \wh\dr_\la[ -\sqrt{-g}T^\la_\al\tau^\al
+\dr^\la_A\cL(u^A{}_\al^\bt\dr_\bt\tau^\al
-u^A{}_\al^{\ve\bt}\dr_{\ve\bt}\tau^\al) - \dr_A\cL u^A{}_\al^\la\tau^\al -
\dr^\m_A\cL\wh\dr_\m(u^A{}_\al^\la)\tau^\al]. \label{K10}
\end{equation}
Let us separate the components of the Euler-Lagrange operator
$$
\cE_L = (\cE_{\al\bt}dg^{\al\bt} +\cE_\al{}^{\g\m}dk^\al{}_{\g\m})\w\om
$$
in the expression (\ref{K10}). We get
\be
&& 0\ap \wh\dr_\la[-2g^{\la\m}\tau^\al\cE_{\al\m} -u^A{}_\al^\la\tau^\al
\cE_A] +\\
&& \qquad \wh\dr_\la[ \dr^\la_A\cL u^A{}_\al^\m\dr_\m\tau^\al
 - \wh\dr_\m(\dr^\m_A\cL u^A{}_\al^\la)\tau^\al
+\wh\dr_\m(\pi_\al{}^{\ve\m\la})\dr_\ve\tau^\al] +\\
&& \qquad \wh\dr_\la[-\wh\dr_\m(\pi_\al{}^{\nu\m\la}\dr_\nu\tau^\al)]
\ee
and then
\be
&& 0\ap \wh\dr_\la[-2g^{\la\m}\tau^\al\cE_{\al\m}
-(k^\la{}_{\g\m}\cE_\al{}^{\g\m} - k^\si{}_{\al\m}\cE_\si{}^{\la\m} -
k^\si{}_{\g\al}\cE_\si{}^{\g\la})\tau^\al +
\cE_\al{}^{\ve\la}\dr_\ve\tau^\al]
+\\
&& \qquad \wh\dr_\la[
 - \wh\dr_\m(\dr_\al{}^{\la\m}\cL)\tau^\al]
 +\\
&& \qquad
\wh\dr_\la[-\wh\dr_\m(\pi_\al{}^{\nu\m\la}(D_\nu\tau^\al+
\Om^\al{}_{\nu\si}\tau^\si)].
 \ee
The final form of the conservation law (\ref{K8}) is
\ben
&& 0\ap \wh\dr_\la[-2g^{\la\m}\tau^\al\cE_{\al\m}
-(k^\la{}_{\g\m}\cE_\al{}^{\g\m} - k^\si{}_{\al\m}\cE_\si{}^{\la\m} -
k^\si{}_{\g\al}\cE_\si{}^{\g\la})\tau^\al +
\cE_\al{}^{\ve\la}\dr_\ve\tau^\al- \wh\dr_\m(\cE_\al{}^{\la\m})\tau^\al]
+ \nonumber\\
&& \qquad
\wh\dr_\la[-\wh\dr_\m(\pi_\al{}^{\nu\m\la}(D_\nu\tau^\al+
\Om^\al{}_{\nu\si}\tau^\si)]. \label{K11}
 \een

It follows that the SEM conservation law in the affine-metric gravity is
reduced to the form (\ref{K1}) where $U$ is the generalized
Komar superpotential (\ref{K2}).

Note that, as like as in gauge theory, the fact that this superpotential
depends on the components of a vector field $\tau$ provides invariance of the
SEM conservation law (\ref{K1}) under general covariant transformations.

Let us now consider the total system consisting of the affine-metric gravity
and  tensor fields described above, e.g., a Proca field. In the presence of a
general linear connection, their Lagrangian density $L_m$ is naturally
generalized through covariant derivatives and depends of the torsion.
The total SEM conservation law is the sum of the expression  (\ref{C313}) and
(\ref{K11}) plus the additional contribution  $$ \wh\dr_\m
(\dr_\al{}^{\la\m}\cL_m\tau^\al) $$
in the superpotential term.

One can consider general linear connections in the presence of a background
world metric $g$
when the general covariant transformations are not exact. In this case, the SEM
complex of affine-metric gravity takes the form
$$
T^\la = \sqrt{-g}T^\la_\al\tau^\al +
\wh\dr_\m(\pi_\al{}^{\nu\m\la}(D_\nu\tau^\al+
\Om^\al{}_{\nu\si}\tau^\si))
$$
and the conservation law comes to the form of the familiar covariant
conservation law
$$
 T^\la_{\m;\la}\ap 0.
$$
\bigskip

\centerline{\bf Acknowledgement}
\medskip

The authors thank Prof.L.Mangiarotti for fruitful discussion.

\end{document}